\documentclass[reprint,pre,showpacs,showkeys,floatfix]{revtex4-1}
\usepackage[T1]{fontenc}
\usepackage{lmodern}
\usepackage{textcomp}
\usepackage[cp1250]{inputenc}
\usepackage[english]{babel}
\usepackage{graphicx}
\usepackage{color}

\begin{document}

\title{Role of detritus in a spatial food web model with diffusion}
\author{Andrzej P\k{e}kalski}
\email{andrzej.pekalski@ift.uni.wroc.pl}
\affiliation{Insitute of Theoretical Physics, University of Wroc{\l}aw, 
pl. M. Borna 9, 50-254 Wroc{\l}aw, Poland}
\author{Janusz Szwabi\'{n}ski}
\email{janusz.szwabinski@ift.uni.wroc.pl}
\affiliation{Insitute of Theoretical Physics, University of Wroc{\l}aw, 
pl. M. Borna 9, 50-254 Wroc{\l}aw, Poland}

\pacs{87.10.Mn, 87.10.Rt, 05.40.-a}
\keywords{Monte Carlo simulations; Food webs; Nutrient cycle; Detritus; Diffusion; Spatial dispersion}

\begin{abstract}
One of the central themes in modern ecology is the enduring debate on whether 
there is a relationship 
between the complexity of a biological community and its stability. In this 
paper, we focus on the role of detritus and spatial dispersion on the stability 
of ecosystems. Using Monte Carlo simulations we analyze two three level models 
of food webs: a grazing one with the basal species (i.e. primary producers) 
having unlimited food resources and a detrital one in which the basal species 
uses detritus as a food resource. While the vast majority of theoretical studies 
neglects detritus, from our results it follows that the detrital food web is 
more stable than its grazing counterpart, because the interactions mediated by 
detritus damp out fluctuations in species' densities. Since the detritus model 
is the more complex one in terms of interaction patterns, our results provide 
new evidence for the advocates of the complexity as one of the factors enhancing 
stability of ecosystems.   
\end{abstract}

\maketitle

\section{Introduction}

The concept of a food web (FW) as a network of local trophic interactions 
between species dates back 
to the pioneering work of Elton~\cite{elt27}. 
Since then, properties of food webs have become subject of intensive studies from both an experimental and a theoretical point of view~\cite{mor01,
dro01,ama99,pek08,moo07,ata08,dea89a,fil10a,fil10b}.

Much of the FW research has oriented itself around various aspects of ecosystem 
stability. Until early 1970s the
 predominant view of ecologists was 
that complex communities are more stable than simple ones 
~\cite{mac55,odu53,elt58}. (Complexity is connected here with an 
increase of the number of trophic levels and links among them.) 
This view was then confounded by May~\cite{may72}, who showed 
mathematically by making use of random matrices that complexity tends to destabilize community dynamics.
Most of the subsequent work related to food 
webs was devoted to finding mechanisms that would allow complex communities to persists~\cite{pim78,yod81,til94,til96,sch99,san99,hus97,til97}. 
Recent advances indicate that on average complexity can be expected to enhance stability. 
However, the complexity-stability debate is far from being 
over and further efforts are needed to fully understand the  relationship between them.

A problematic aspect of the debate is the lack of consensus about how ecological stability should be defined. The notion of stability is actually a 
catch-all term that can refer to persistence, resilience, resistance or robustness of a system. In general, the plethora of definitions of ecological 
stability may be divided into two categories - definitions that are based on a system's dynamic stability and definitions that are based on a system'
s ability to defy change~\cite{mcc00}. In the context of conservation ecology, 
stable populations are often defined as ones that do not go extinct. 
We will follow this practice throughout this paper. Problems 
related to the reaction of the system to external perturbations are far more 
complicated, need more parameters and therefore such  studies should be 
carried out after the basic mechanisms are well 
understood~\cite{mol94}.

Several mechanisms have been already identified as factors leading to the stability of food webs: compartmentalization, generalist consumers, top-down control or weak-interaction effects, among others. Compartments, i.e. subsets of species that interact more frequently among themselves than 
with other species, are known to buffer the propagation of extinctions~\cite{the10,sto11}. Generalist consumers, which are able to switch from one 
food source to a more abundant one, keep a food web stable, because they control the abundant species and let the less common one recover~\cite{tho07,
neu07}. Top-down control of lower trophic levels by an apex predator promotes stability as well, because it limits the degree to which prey endanger 
primary producers~\cite{est11}. The structure of a food web itself may influence its stability in many ways. For instance it has been shown that 
species weakly linked with other ones stabilize community dynamics by dampening destabilizing consumer-resource interactions~\cite{odu53,mcc00}. 

Even though recent years have seen significant progress in understanding stability factors in ecosystems, there are at least two mechanisms which 
have been consequently neglected in the vast majority of theoretical studies: detritus and spatial dynamics. Being an energy source and nutrient 
reservoir, detritus (i.e. dead organic matter) plays an important role in nutrient cycling and food web dynamics~\cite{dea89b, dea92, hai93, moo93, 
pol96}. In many ecosystems detrital chains are the major pathway of energy flow~\cite{hai93,pol96} and may have stabilizing effect on trophic 
dynamics~\cite{dea92, moo93, moo04}. However, existing theories on food webs have largely neglected detritus-based systems and have focused on 
grazing food webs~\cite{moo04} with living plant biomass (or net primary production) as the source of energy and nutrient for the first-level 
consumers.

Food webs are spatial entities consisting of species which use dispersal for both avoiding natural enemies and searching for food. There is an 
emergent consensus that spatiotemporal patterns and processes are crucial for their stability~\cite{moo04,sch89,pol96,hol02}. However, until recently 
that consensus was not reflected in ecological theories which usually abstracted away from the dispersal of ecosystems~\cite{ama08}. 

There is already evidence in existing literature suggesting that many questions asked in ecology on stability of food webs should be revisited after 
incorporation of both detritus~\cite{ata08} and dispersal~\cite{hol02}. Thus, any insight into the role of these two factors is valuable to our 
understanding of real ecosystems.

In our recent studies~\cite{szw10,szw12,szw13} a food web model with a detritus path and a spatial dynamics was analysed by means of Monte Carlo 
simulations. The model consists of three trophic levels, each of which is populated by animals of one distinct species. While the species at the 
intermediate level feeds on the basal species, and is eaten by the predators living at the highest level, the basal species itself uses the detritus 
as a food resource. Our results are consistent with the hypothesis of top-down interactions being essential for the stability of food webs~\cite{
est11}.

We have found that in certain conditions complex spatiotemporal patterns in the form of density waves appear in the model. These waves travel through the entire system and drive it to extinction. We have shown that the waves are triggered by the spatial accumulation of detritus.

The model presented in~\cite{szw10,szw12,szw13} has one important shortcoming: all individuals remain localized, i.e. once put on the lattice they do 
not move. The only way of invading new lattice sites is via proliferation. While it may be reasonable for the basal species, the assumption is rather 
unrealistic for the animals populating higher trophic levels and may impeach  
the generality of the results.  The aim of the present study is thus to relax this assumption and to incorporate diffusion of animals into the rules 
governing the dynamics of the model.   To investigate the feedback between detritus and stability, we will compare the model with a three level 
grazing food web, in which detritus is omitted and the basal species has access to unlimited food resources. 

Apart from mobility of animals, the present model differs from the one considered in ~\cite{szw10,szw12,szw13} in several other aspects. Organisms 
forming a trophic level have now individual characteristics, their fate (eating, proliferation) is determined by local conditions and they may die of 
hunger if they do not find food fast enough.

The paper is organized as follows. In Section 2 the model is briefly introduced. Simulation results are discussed in Section 3. And
finally in Section 4 conclusions are drawn.

\section{Model}

\begin{figure}
\centering
\includegraphics[scale=0.5]{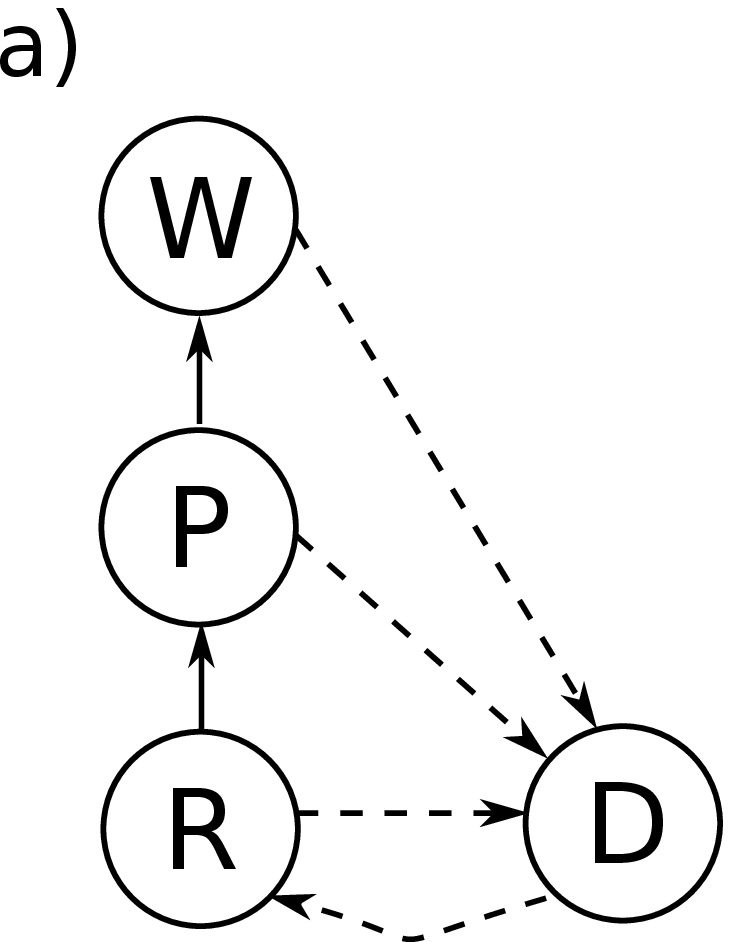}\ \hspace{2cm}
\includegraphics[scale=0.5]{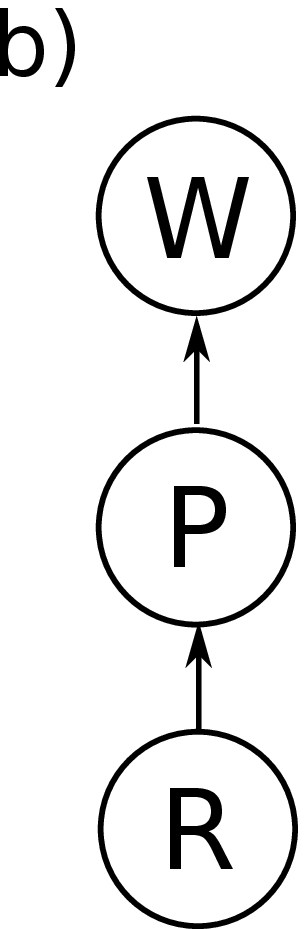}\
\caption{Models investigated in this paper: the detritus food web (left) and the grazing one (right). The arrows indicate the flow of nutrient in the 
food webs (with dashed lines being related to the dead organic matter). In both cases, prey $P$ feed on the primary producers $R$ and are themselves 
food for predators $W$. The primary producers feed on detritus $D$ (left) or have unlimited food resources (right).\label{food web models}}
\end{figure}

Fig.~\ref{food web models} (left panel) illustrates the model we are going to investigate. It consists of three trophic levels, each of which is 
populated by a distinct taxon. The basal level species corresponds to primary producers in real ecosystems. It will be denoted henceforth as $R$. The 
species at the intermediate level, i.e. prey $P$, relates to herbivores, which feed on primary producers. The prey themselves constitute food for the 
top level species - predators $W$. The predators correspond to carnivores in real systems. The remains of individuals form detritus $D$, which 
provides nutrient for the primary producers. For the sake of simplicity we will assume that the conversion of dead fragments into nutrient occurs 
immediately and without any external help. However, in reality dead organisms are broken down and turned into useful chemical products by decomposers.
 We will address their impact on food web stability in a forthcoming paper. 
 
Our aim is to investigate the model by means of Monte Carlo simulations. All individuals are put on a square lattice with hard wall boundary 
conditions, meaning that any action taking an animal outside the lattice is prohibited. While the primary producers remain localized on the lattice, 
prey and predators can move to neighboring sites. However, double occupancy of nodes by agents of the same type is not allowed.

Like in real ecosystems, each individual in our model must feed in order to survive. It stores energy derived from food and then consumes it to maintain its activities. It is characterized by a metabolic rate, i.e. a parameter which specifies the amount of energy needed to survive one simulation step without feeding. In real systems metabolic rates vary from one individual to the next depending on factors such as age, sex, body composition and physical activity level. However, in our simple model we do not differentiate between organisms of a given species and assume that all agents of the species $i$ ($i=R,P,W$) have exactly the same metabolic rate $m_i\in (0,1)$. The energy reserve of an individual is replenished back to its maximum level (equal to 1 for all agents in the simulation) after each feeding and decreased by $m_i$ after each Monte Carlo step (MCS), which is our time unit. Thus, if an agent was not able to find food in $1/m_i$ steps, it dies from starvation and is turned into detritus. A model of animals having energy reserves, which have to be refilled in order to survive, has been already introduced to describe effects of predation in a three species model~\cite{szw06}.

Once put randomly on the lattice, the system will evolve due to the following rules:
\begin{enumerate}
\item Pick a lattice site at random.
\item If the chosen site is empty or occupied by only a $D$, do nothing.
\item If the site is occupied by a $W$ or a $P$, move the agent to a site in the Von Neumann neighborhood, provided it is not occupied by an agent of 
the same kind. If there is food at the new site, increase the energy reserve of the agent, then transform its prey into detritus and finally produce 
offspring, i.e. put a progeny at an appropriate site in the  Von Neumann neighborhood with half of the maximum value of the 
reserves for its species. Appropriate means here not occupied by an agent of the same type or a potential predator.
\item If the site is occupied by an $R$ and if there is $D$ at that site, increase the energy reserve of the primary producer, remove $D$ from the 
lattice and  produce offspring (see step 3 for more details).
\item Repeat steps 1.-4. $N^\prime$ times to complete one Monte Carlo step. Here, $N^\prime$ is the number of all agents on the lattice at the 
beginning of the step. An agent can be chosen just once in a given MCS.
\item At the end of each time step, reduce the energy reserves of all agents by $m_i$. Then, convert agents with the reserve equal to 0 into detritus (death 
due to starvation). We decided that an occurrence of a $D$ is more important than its actual quantity. As a consequence, if  a $P$ or  a $W$ dies at a node already containing a $D$, then its content will remain one $D$.

\item Repeat the whole procedure a given number of times.   
\end{enumerate}

For the sake of comparison, we will also consider a grazing version of the model (right panel in Fig.~\ref{food web models}) with no detritus and 
with primary producers having unlimited food resources. In this case the basal species cannot die due to the lack of food. All other characteristics 
of the food web remain the same.

Parameters of the model are the following:
\begin{enumerate}
 \item Lattice size. We took it as $L$ = 100. Increasing $L$  slows down
the dynamics without affecting the results.
\item Initial concentration of species. We took $W(0)$ = 0.2 and 0.4 for the
remaining species. Reducing it by half leads to less rich dynamics with coexistence region extending  
to very low metabolic rates. We have however studied the role of 
the initial concentration of predators, see below.
\item Metabolic rates - $m_W, m_P, m_R$. These are our control parameters.
\item Number of offspring. We took 1 for predators, 2 for prey and 4 for producers.
This assumption complies with the fact that the lower the trophic
level the higher its productivity~\cite{odu57}. 
Since a progeny might be put only on a plaquette
without an animal of the same type or a potential predator (otherwise it is not born),
increasing these numbers would not change the results in any significant
way. 
\end{enumerate}

The above assumptions concerning the counting of detritus and the number of offspring at each trophic level imply that the model constitutes an open system with the total biomass being not conserved. Despite the fact that many real ecosystems are open as well~\cite{mat07} we understand our model as one of possible food web motifs (i.e. a subgraph of a food web)~\cite{cam07,mcc98} rather than a representation of an existing ecosystem. Since motifs are usually connected with other motifs, a biomass flow between them is possible even if the whole ecosystem is closed. 

\section{Results}

All results presented in this section were obtained in Monte Carlo simulations with the Von Neumann neighborhood (i.e. 4 neighboring sites) on the 
square lattice. Other choices are possible, but they are more computationally demanding and do not qualitatively affect the results.

Most simulations were performed up to 1000 MCS, because at that time the system has usually reached the stationary state. In other words, it is 
highly unlikely that a system which   is alive at $T=1000$ MCS, will die afterwards. The stationarity of the time series has been checked with two 
complementary tests, the Augmented Dickey-Fuller test~\cite{sai84} and the Kwiatkowski-Phillips-Schmidt-Shin one~\cite{kwi92}, both implemented in 
the R package \textit{tseries}~\cite{tra13}.

If required (e.g. for survival chance diagrams discussed in Sec.~\ref{sec survival chance}), we averaged the results over 50 independent runs. This 
particular number of runs yields already a reasonable statistics. However, to analyze the time evolution of the model, we will often look at the 
results of single runs, because averaging could smooth out the effects of fluctuations which are crucial for the survival of the system.   

\subsection{Time evolution of the model food web}
\label{sec time evo}

\begin{figure*}
 \centering
 \includegraphics[scale=1]{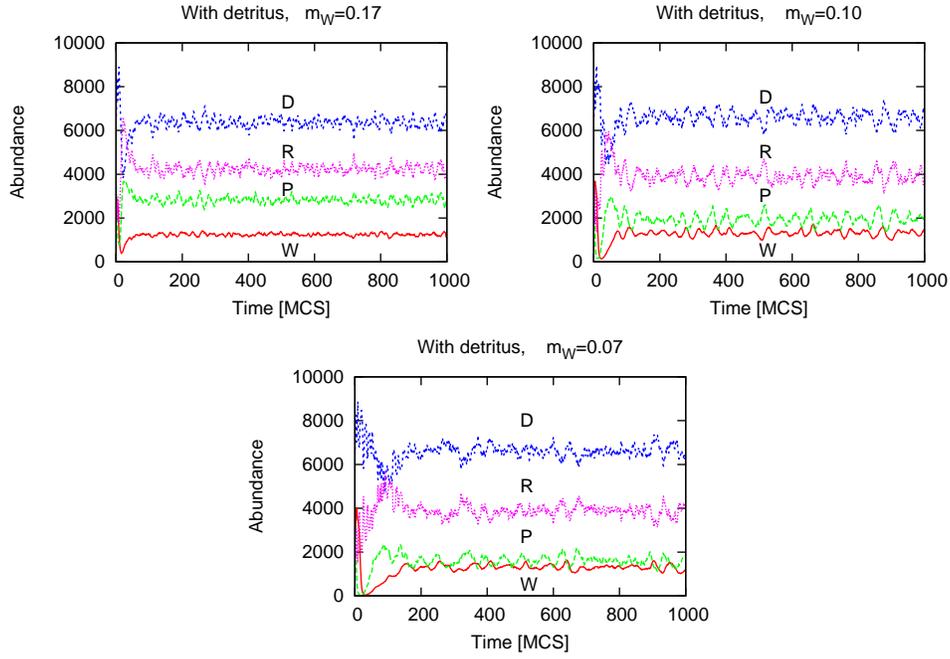}
 \caption{(Color online) Abundance of predators $W$, prey $P$, producers $R$ and detritus $D$ in the model with detritus for different values of $m_
W$. Other metabolic rates were fixed throughout the simulations ($m_R=m_P=0.1$). Results from single runs. 
\label{abundance_det}}
\end{figure*}

\begin{figure*}
 \centering
 \includegraphics[scale=1]{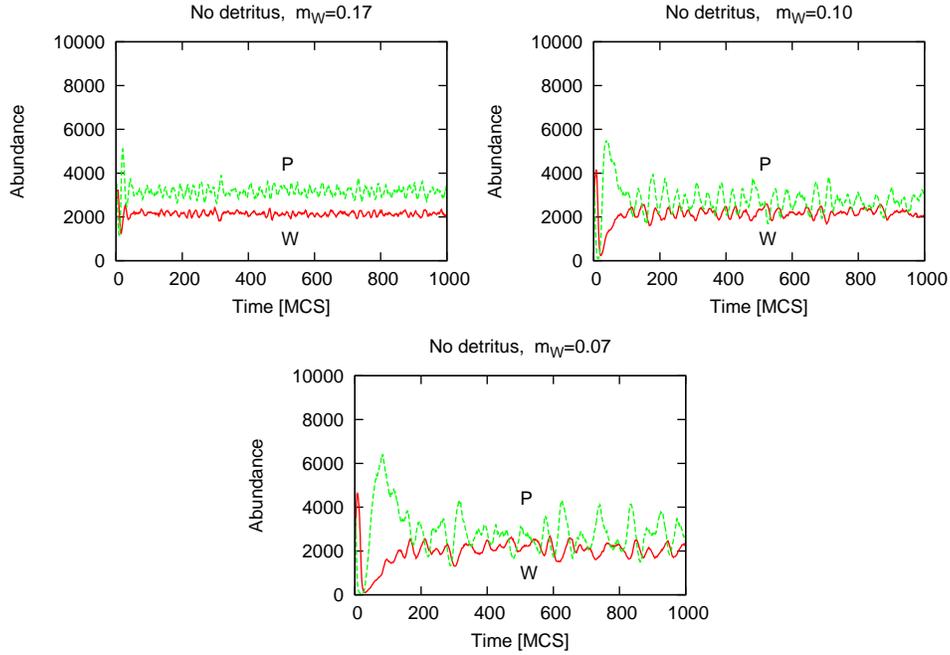}
 \caption{(Color online) Abundance of predators $W$ and prey $P$ in the model without detritus for different values of $m_W$. Other metabolic rates 
were fixed throughout the simulations ($m_R=m_P=0.1$). Results from single runs. 
\label{abundance_graz}}
\end{figure*}

We start our analysis with the time evolution of population densities (see
Fig.~\ref{abundance_det} for the detritus model and Fig.~\ref{abundance_graz} for the
grazing one).
The plots have been obtained for different values of the predators' metabolic rate ($m_W=0.17,0.1,0.07$). The rates of other species were fixed ($m_
R=m_P=0.1$). First of all we see that in the model with detritus changing the metabolic rate of predators has only a small effect on the system. The 
abundance of prey shifts slightly towards smaller values and the amplitude of fluctuations increases with decreasing $m_W$. 

At first glance, there is no big difference between the detritus model and the grazing one. The effect of changing $m_W$ in the food web without 
detritus is qualitatively the same: the number of prey goes down and the amplitude of fluctuations increases with decreasing $m_W$. However, a closer 
look at Figs.~\ref{abundance_det}-\ref{abundance_graz} reveals two important differences. First, there are on average more predators and less prey in 
the grazing model. Second, the fluctuations are much larger in the absence of detritus. Hence we expect the grazing food web to be more fragile, 
because under certain circumstances strong fluctuations may lead to overhunting and the extinction of both species: the prey will be wiped out by the 
numerous predators, which then go extinct due to the lack of food. 

In our simulations we have indeed observed, that the grazing food web is very sensitive to random fluctuations at low values of $m_W$. While the 
results obtained in distinct runs were actually the same in the model with detritus, changing only the seed of the random number generator could 
cause the switch from a stationary coexistence state (all species alive) to the absorbing one with $P$ and $W$ being extinct in the model without detritus. This result indicates that detritus has a stabilizing effect on food webs, in agreement with~\cite{dea92,moo93,moo04}.

Keeping $m_W$ fixed and varying $m_P$ instead yields results similar to Figs.~\ref{abundance_det}-\ref{abundance_graz}, with even smaller dependence 
of the abundances and their fluctuations on the metabolic rate. This could be in turn interpreted as an indication that predators are the key species 
in the food web~\cite{est11}, because changing their characteristics slightly may have a severe impact on the entire system.  
\begin{figure*}
 \centering
 \includegraphics[scale=1]{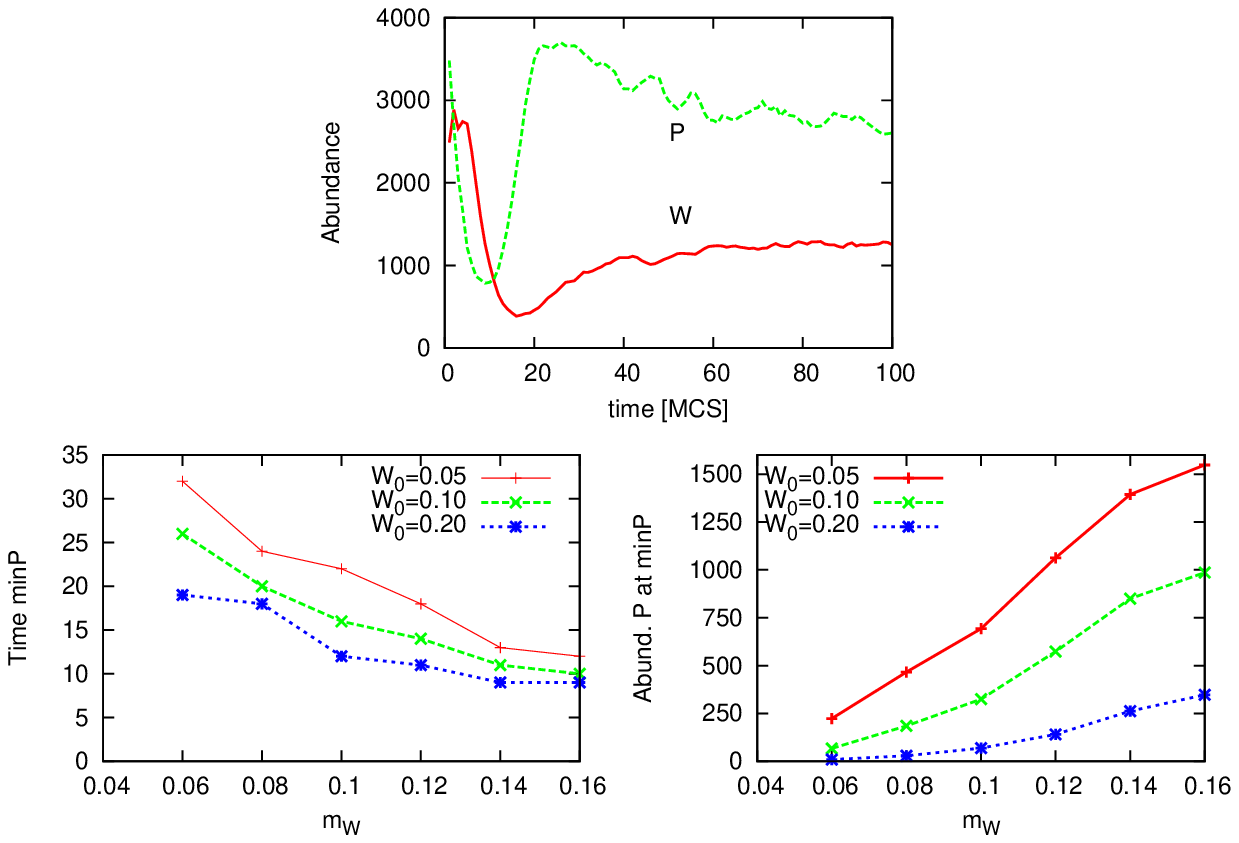}
 \caption{(Color online) Abundance of predators $W$ and prey $P$ in the model with
detritus for $m_W = 0.17$. Fragment of the figure \ref{abundance_det}. (top diagram).
Location of the minima (left bottom diagram) for the abundance of $P$ and their
values at the minimum (right bottom diagram), as functions of predators metabolic
rates and for three values of the initial concentration of predators. $W_0$ is the initial concentration of predators.
\label{minima}}
\end{figure*}
In Fig.~\ref{minima} we show the initial stage of the evolution
shown in Fig. \ref{abundance_det} in the top left diagram. As could be seen, there
is a sharp minimum in the abundance of $P$ and a shallower one
for $W$. The same
type of behavior is found for other values of the parameters. The question one may
ask is what is the influence of the initial number of predators and their metabolic
rate on the location and depth of the producers' minima. Figure \ref{minima} shows
that the differences have rather quantitative than qualitative character. Predators
with low metabolic rates are more effective, i.e. they leave a smaller number of $P$
at the minimum and that minimum happens later than when the predators have higher
metabolic rates and have to kill faster in order to survive. 

\subsection{Survival chance}
\label{sec survival chance}

\begin{figure*}
 \centering
 \includegraphics[scale=0.5]{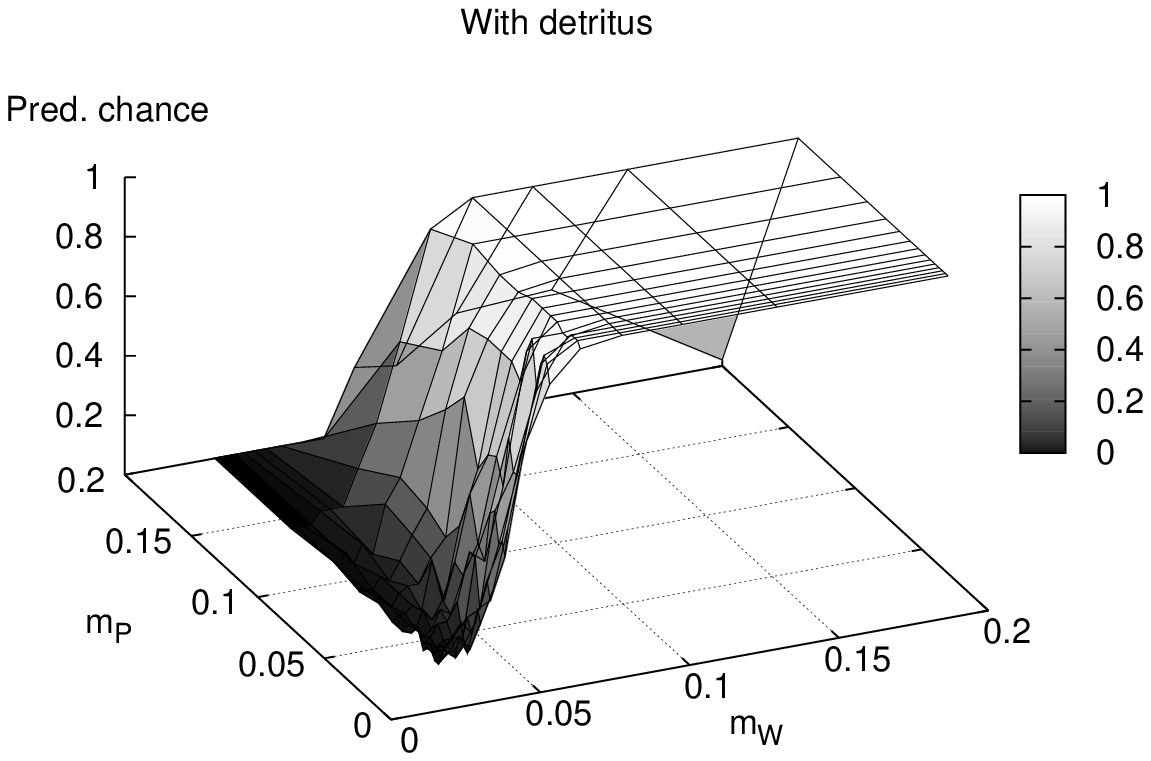}
 \includegraphics[scale=0.5]{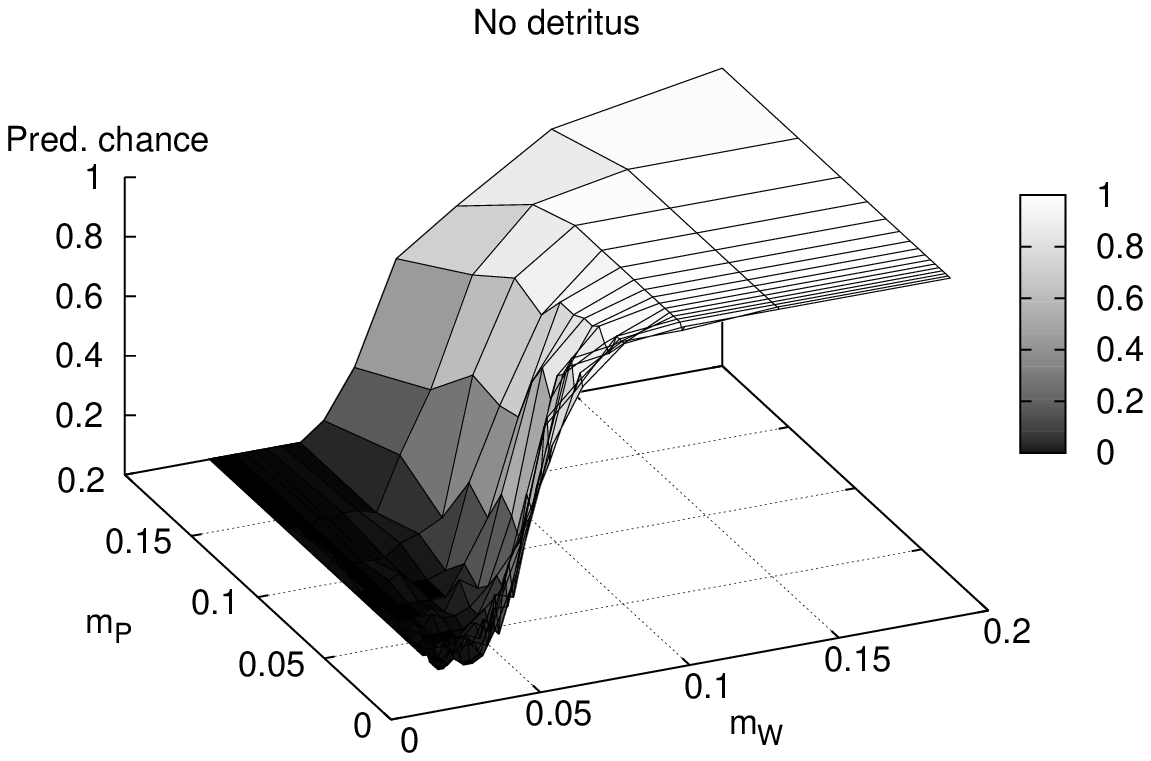}
 \includegraphics[scale=0.5]{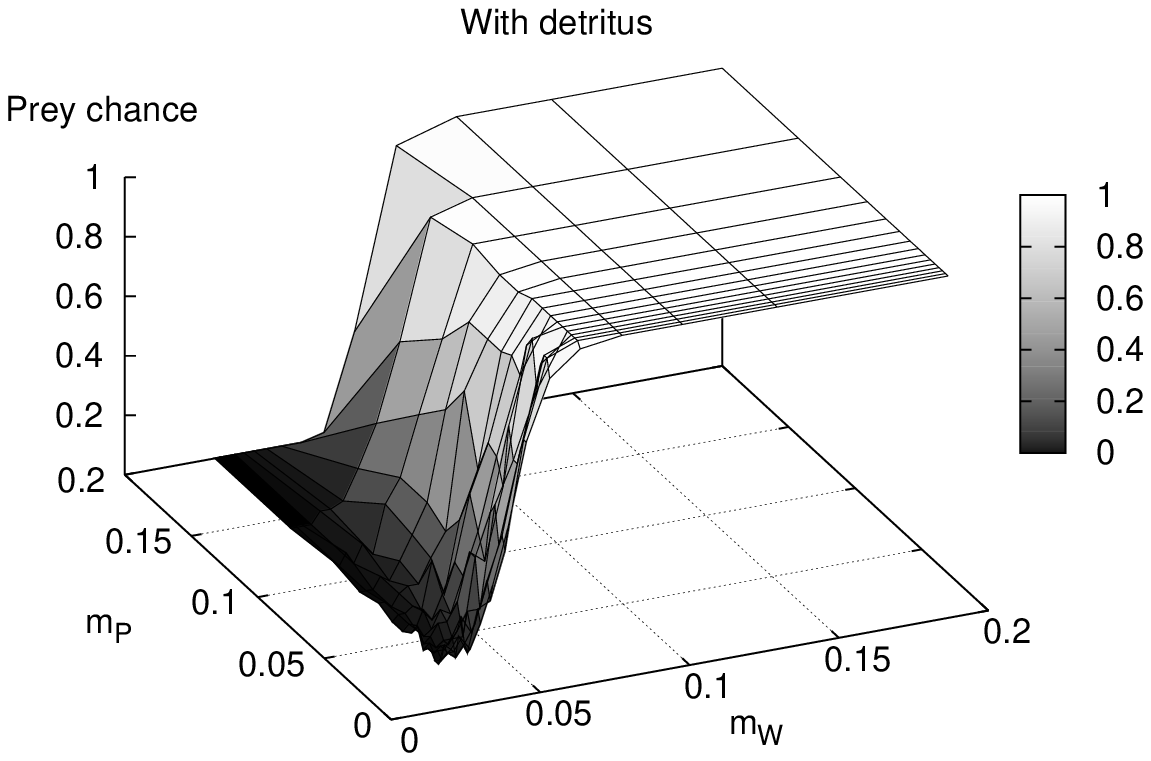}
 \includegraphics[scale=0.5]{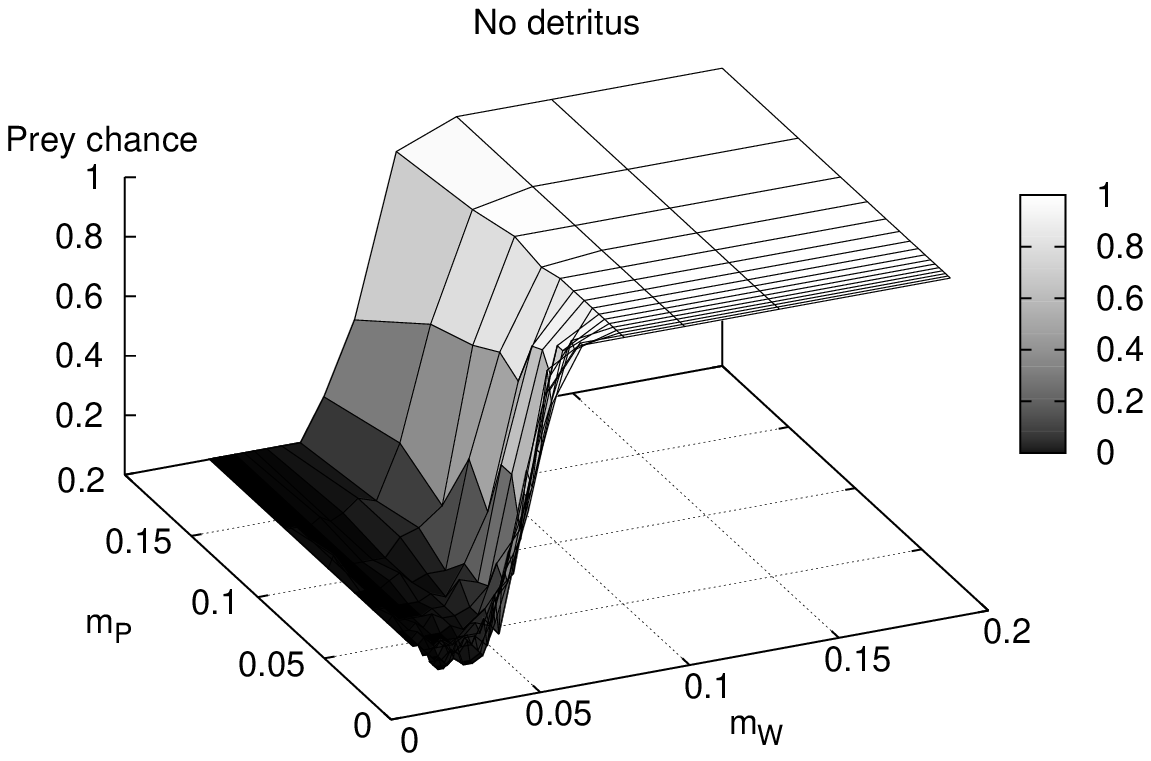}
 \caption{Survival chance of predators  and prey in the detrital (left column) and the grazing food web (right column). Since both models turned out 
to be rather insensitive to changes of $m_R$, we kept the metabolic rate of the primary producers fixed ($m_R=0.1$).\label{survival}}
\end{figure*}

To elaborate on the differences between the models let us investigate chances for a species to survive till the end of simulations. To this end for a 
given set of the control parameters we counted the number of cases among the independent runs when a population survives. Since both models turned 
out to be rather insensitive to changes of $m_R$, we kept the metabolic rate of the primary producers fixed ($m_R=0.1$). The results for the 
population of the prey $P$ and the predators $W$ are shown in Fig.~\ref{survival}.  
The plots may be interpreted as phase diagrams of the system in the $(m_P,m_W)$ plane with the alive phase corresponding to non-zero values of the 
survival chance and the dead phase in which the survival chance is equal to zero. We see that there exists a critical value of $m_W$ below which the 
system is driven to extinction. It is due to the fact that animals with low metabolic rates (long resource consumption) are very effective hunters (
they have 
more time to find even a distant prey) and may eat up all the prey and then die due to the lack of food.

It is important to note that there is no similar critical value of $m_P$. Varying the metabolic rate of prey has only a little effect on the survival 
chance of the system. This finding confirms the hypothesis about top level predators being the key species in food webs~\cite{est11}.

\begin{figure}
\centering
\includegraphics[scale=0.5]{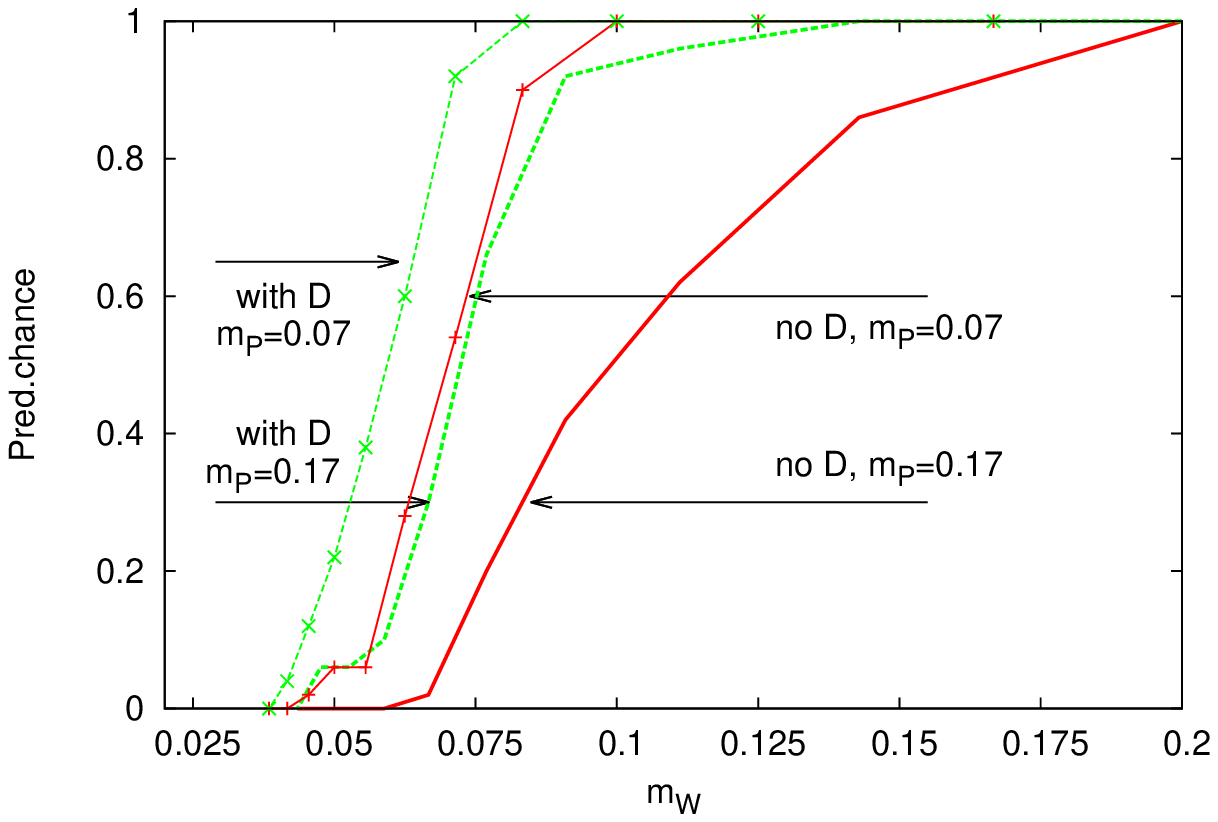}\\
\includegraphics[scale=0.5]{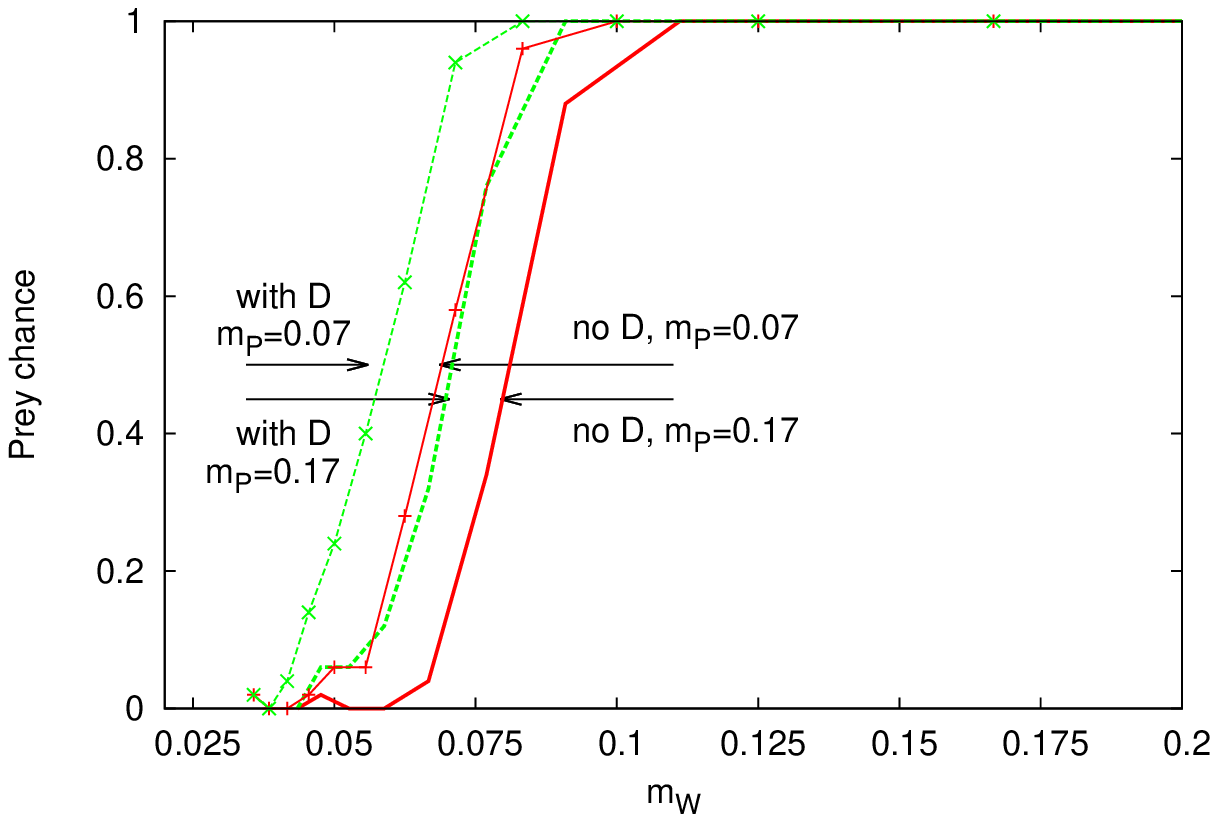}
\caption{ (Color online) Comparison of survival chances of the predators (top panel)
and prey (bottom panel) between the models. Different lines correspond to vertical cross-sections of the surfaces from~Figs.~\ref{survival} for  $m_P$ fixed at different values.
\label{survival_comp}}
\end{figure}

The results are similar for both models, however the critical values differ  from each other. To compare the models in a more convenient way, 
let us look at the 2D projections of the phase diagrams. In Fig.~\ref{survival_comp}
 the survival chance of the predators (top plot)
and the prey (bottom plot) are shown. 
In the model with detritus the populations may survive at lower values of $m_W$. It is evident that the detrital food web is more stable than the 
grazing one, because it is able to sustain more effective hunting being one of the main forces that drive food webs to extinction. 

The detritus food web is of course the more complex one, because the flow of nutrient and the absence of unlimited food resources introduces 
additional implicit interactions between species. These interactions seem to damp the fluctuations of the abundances and in this way enhance the 
stability of the system.
 
This is the most important result of the present paper. It shows that the predictive power of theories on food webs may be questioned if they do not 
include the detritus path. Thus, one has to revisit them before drawing conclusions and applying them to real ecosystems.

\subsection{More on interspecific interactions}

One of the signatures of the interactions in a food web are correlations between the time series describing the time evolution of distinct species. 

\begin{figure*}
 \centering
 \includegraphics[scale=1]{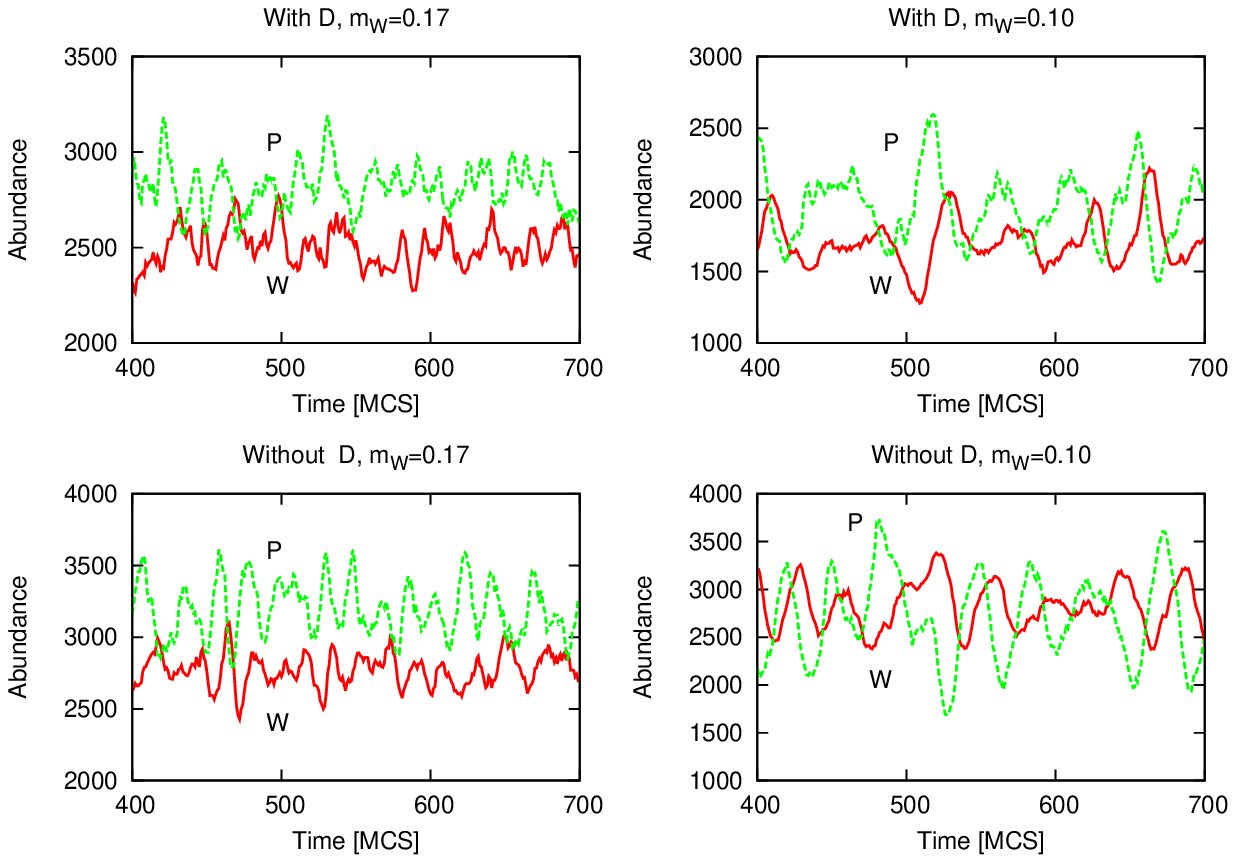}
 \caption{(Color online) Time dependence of the abundances of predators and prey   in a narrow time window for both the detrital FW (top row) and the 
grazing one (bottom row). Curves for $W$ are shifted upwards for the sake of comparison. \label{abundance_narrow_1}}
\end{figure*}

\begin{figure*}
 \centering
 \includegraphics[scale=1]{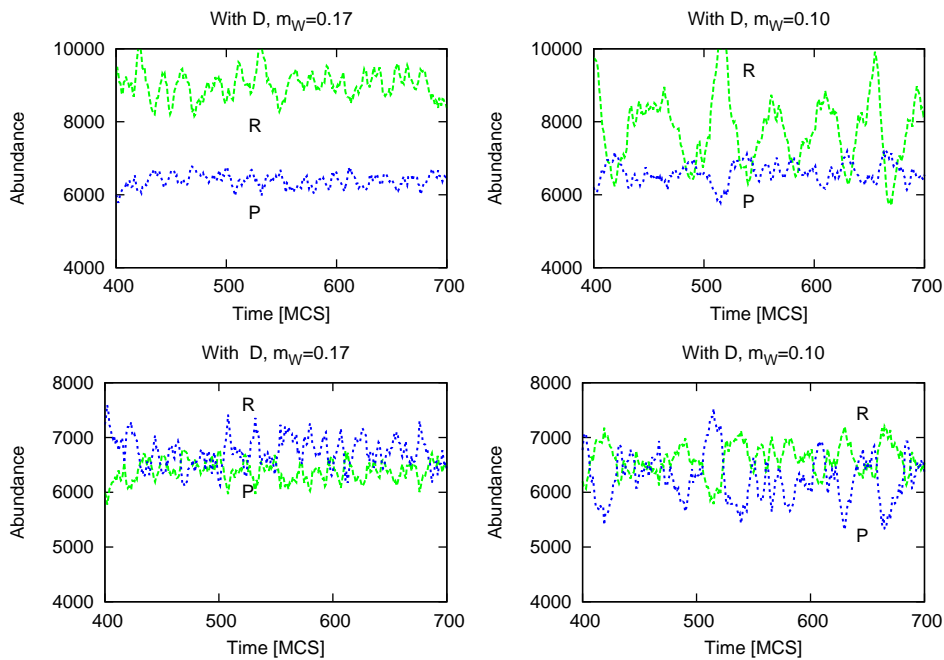}
 \caption{(Color online) Time dependence of  the abundances of prey  and resources in a narrow time window. Curves for $P$ are shifted upwards for 
the sake of comparison. FW with detritus.
\label{abundance_narrow_2}}
\end{figure*}

In Figs.~\ref{abundance_narrow_1}-\ref{abundance_narrow_2} the time evolution of the  models in a narrow time window is shown. To simplify the 
analysis some curves in the plots have been shifted. The plots suggest that the series are correlated somehow, however the variations of the time 
series are too complex to visually judge whether one series tends to lead or lag the other.

To quantify the correlations, we have calculated both the auto- (ACF) and the cross-correlation functions (XCF) for different pairs of the time 
series (refer to~\cite{orf96} for definitions).

\begin{figure*}
 \centering
 \includegraphics[scale=1]{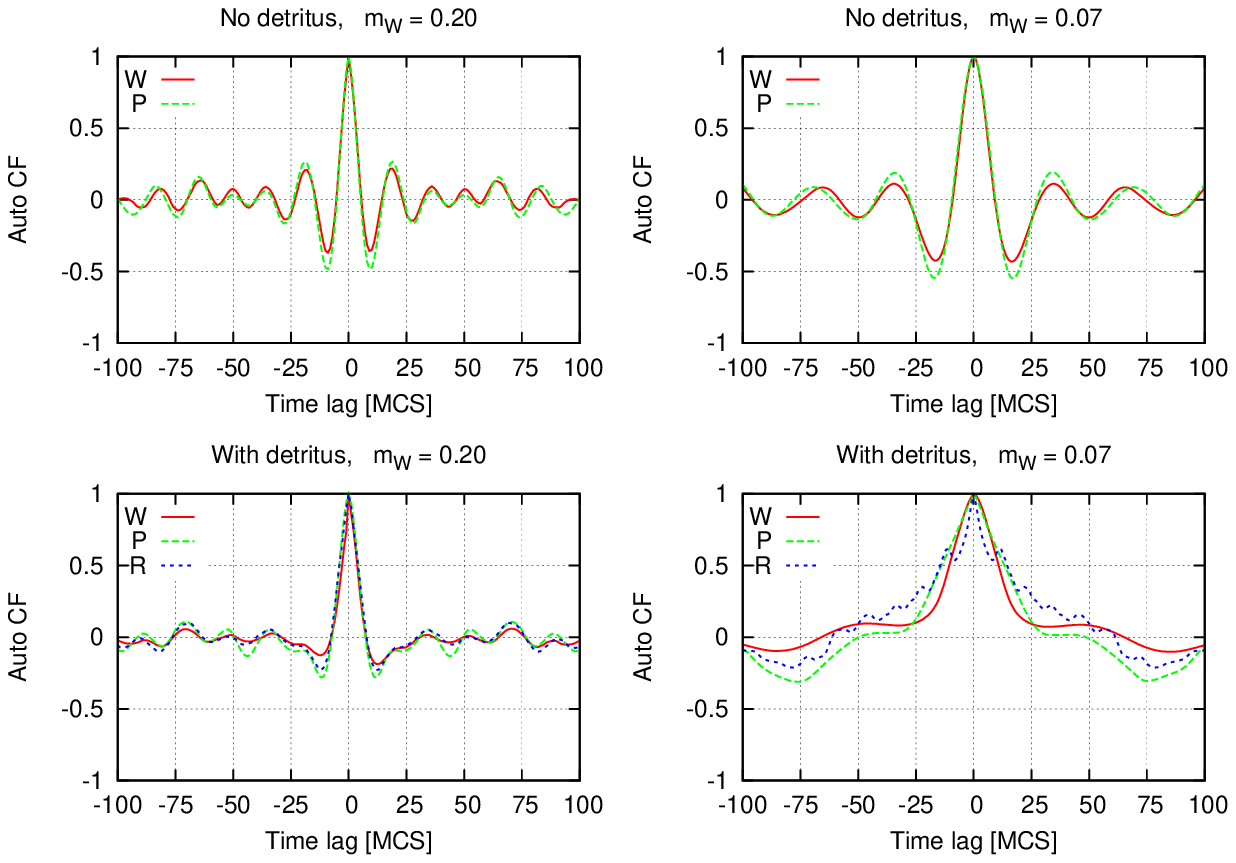}
 \caption{(Color online) Autocorrelation functions for the grazing FW (top row) and for the detrital one (bottom). \label{acf}}
\end{figure*}

The results for the autocorrelation function are presented in Fig.~\ref{acf} for two values of the predators' metabolic rate. In the model without 
detritus (top row in Fig.~\ref{acf}) the ACFs are periodic and show features common to Lotka-Volterra predator-prey systems~\cite{pro99}. This 
should not be very surprising, because in this case the primary producers
populate the whole lattice most of the time. Thus the model is
essentialy a predator-prey system with unlimited food resources for prey, i.e. one of the Lotka-Volterra type.

The picture for the model with the detritus is different. The ACF reveals a broad peak at zero time lag and falls down quickly. The interactions 
mediated by the detritus seem to damp out regular oscillations, which were present in the grazing FW. Note that in this case the ACF changes a lot 
with decreasing $m_W$. The leading peak becomes very wide and the oscillations vanish almost entirely.

\begin{figure*}
 \centering
 \includegraphics[scale=1]{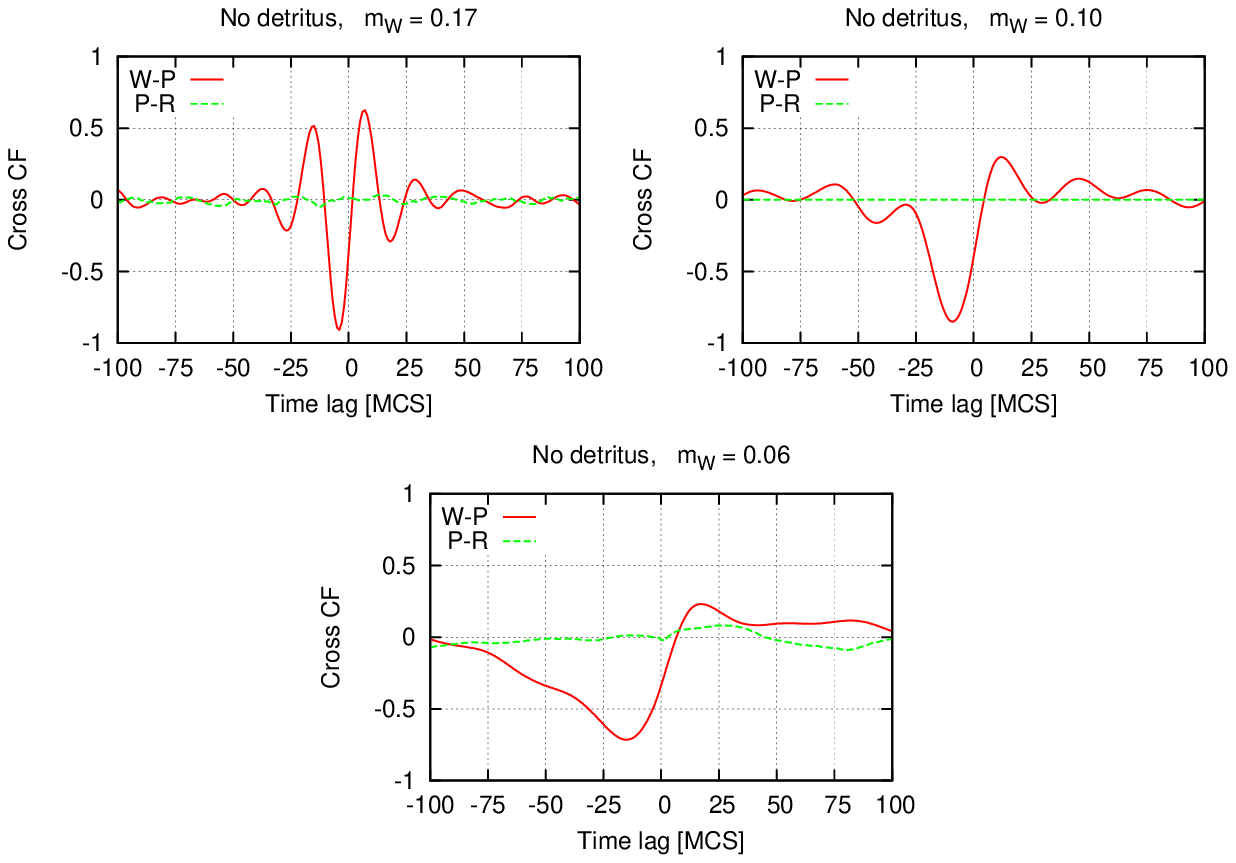}
 \caption{(Color online) Cross-correlation functions for a FW without detritus.\label{ccf_graz}}
\end{figure*}

The cross-correlation functions for the grazing model are shown in Fig.~\ref{ccf_graz}.
We see that there are practically no correlations between the primary producers $R$ and the prey $P$. Again, this result indicates that the grazing 
food web may be treated as a Lotka-Volterra-like system. As far as the correlations between $W$ and $P$ is concerned, we observe a strong positive 
one at the time lag $\tau\simeq 5$, i.e. the maxima in the abundance of the prey precede the maxima in the abundance of the predators. It takes 
only 5 MCS for the predators to adjust to the increasing number of prey. The delay increases with decreasing $m_W$. The reason is rather intuitive: 
predators with a low metabolic rate live longer without food. On one hand, they have
time to search for food and are not any more confined to local prey
clusters  in order to survive, on the other -  looking for food takes longer as well.
That is why the population needs more time to follow
the changes in the abundance of prey.

\begin{figure*}
 \centering
 \includegraphics[scale=1]{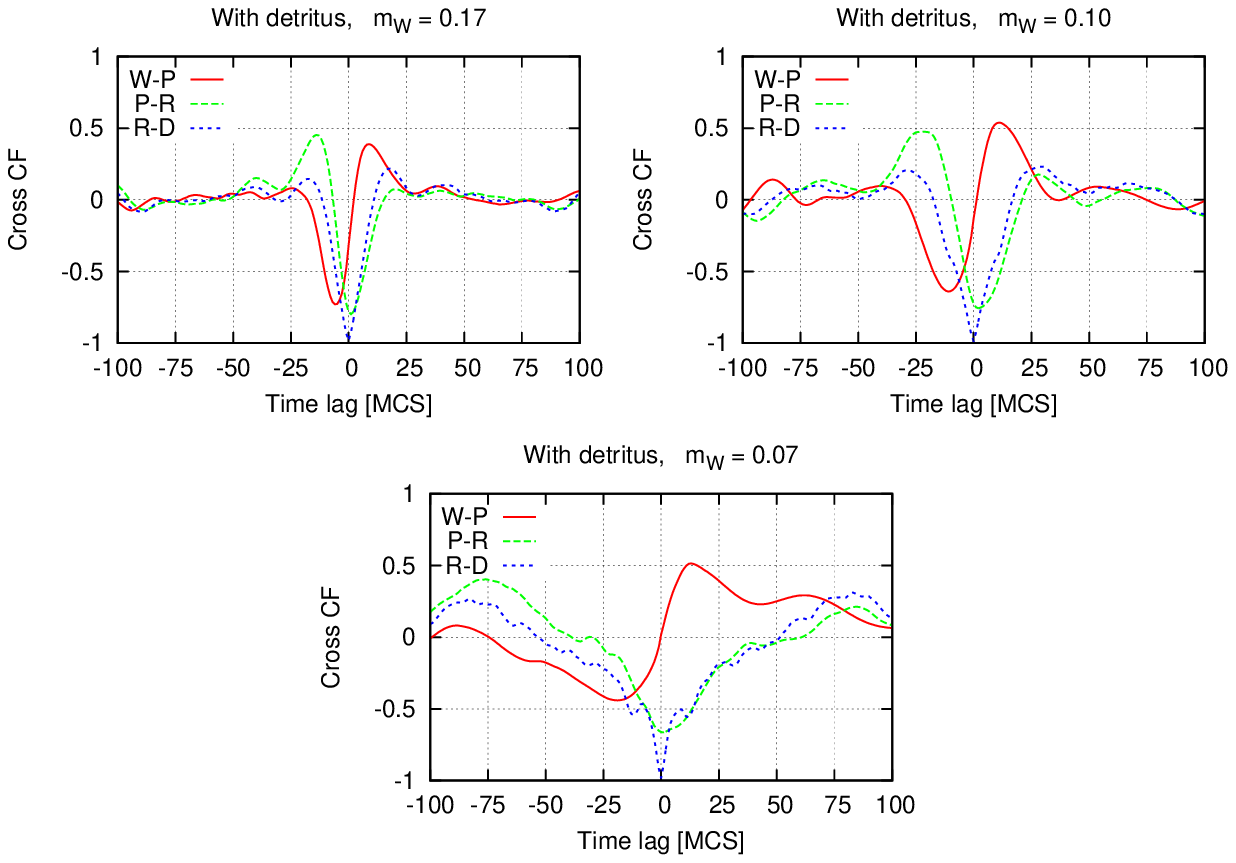}
 \caption{(Color online) Cross-correlation functions for a FW with detritus. 
\label{ccf_det}}
\end{figure*}

The cross-correlations for the detrital food web are shown in Fig.~\ref{ccf_det}.
Due to more complex interaction patterns the picture was expected to be more complicated than in the other model. As can be seen, the primary 
producers $R$ and the detritus $D$ are strongly anticorrelated at zero time lag. That means that the producers adjust instantaneously to the 
accumulation of nutrient. It is so because they are localized and do not move on the lattice. They do not look for food, but discover 
immediately if the food is available or not and act accordingly.

The correlations between the predators $W$ and the prey $P$ are similar to the grazing case. The predators follow the changes in the prey population 
with a delay, which increases with decreasing $m_W$. 

The relationship between the prey $P$ and the primary producers $R$ in Fig.~\ref{ccf_det} 
is very unintuitive. Actually, one could expect here a correlation typical for the predator-prey interaction, i.e. something similar to the $W$-$P$ 
curve with a positive time lag. However, we observe a strong anticorrelation, i.e. the maxima of $P$ are preceded by the minima of $R$. In other 
words, the primary producers follow the changes of prey and not vice versa. This is one of the manifestations of the complex interactions mediated by 
the detritus.   

\begin{figure*}
 \centering
 \includegraphics[scale=0.5]{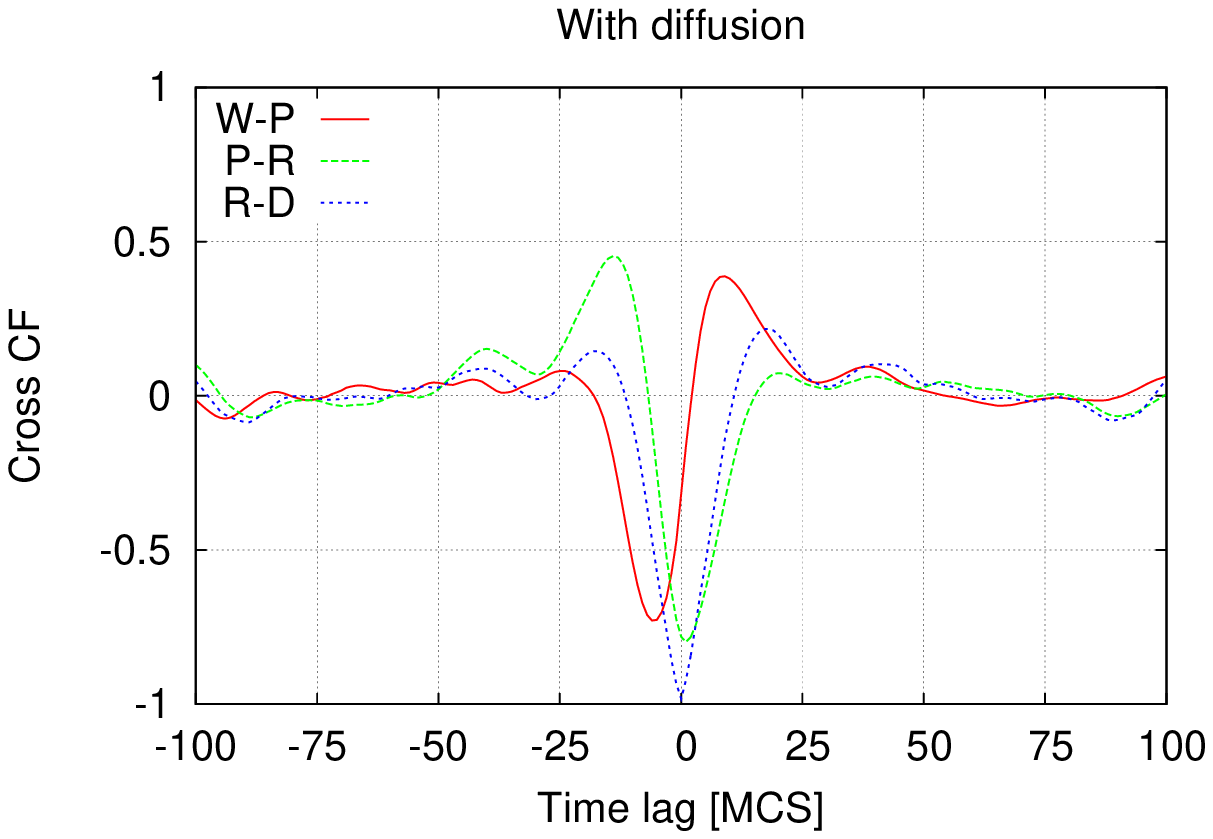}\
 \includegraphics[scale=0.5]{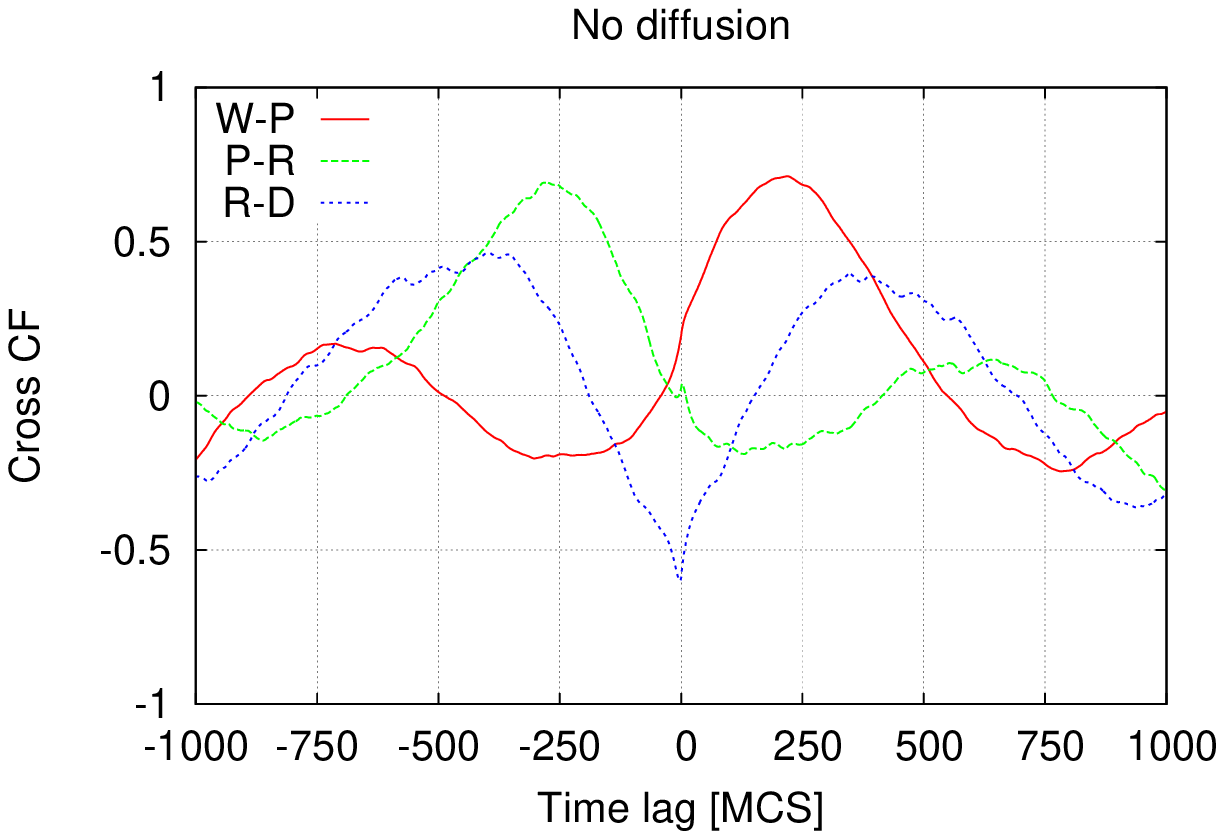}\
 \caption{(Color online) Examples of cross-correlation functions for models with and without the diffusion of animals (see \cite{szw10,szw12,szw13} 
for the detailed analysis of the model without diffusion). Notice the difference
between the time scales in the plots.
\label{xcorrcomp}}
\end{figure*}

It is interesting to check how the model with diffusion differs from the static 
one~\cite{szw10,szw12,szw13}. To recall, in the latter model all agents stayed 
localized, i.e. once put on the lattice, they did not move.
 The only way to invade new lattice sites was via proliferation. In 
Fig.~\ref{xcorrcomp}, as an example the cross correlation functions for both 
models are shown. Apart from the fact that in the model without the diffusion 
all processes take much longer, the corresponding curves are qualitatively very 
similar. These results indicate that the diffusion does not change essentially 
the interaction patterns between species. However, the interactions themselves 
are mediated much quicker due to the diffusion. This shortening of the time 
scale, while not really important in an isolated system, may become crucial if 
the food web is exposed to an external perturbation varying in time, because the 
interplay between different time-scale phenomena may impact the 
dynamics~\cite{ben07}. We will address this issue in a forthcoming paper.

\section{Conclusions}

A simple spatial food web model in a detrital and a grazing version has been presented and analyzed by means of Monte Carlo simulations. 

Our results give some new evidence for the advocates of the complexity as one of the factors enhancing stability of ecosystems in the enduring 
complexity-stability debate which is the central theme in the modern ecology.

We have shown that the food web with detritus is more stable than its grazing counterpart. This agrees with the experimental findings~\cite{dea89a,
moo93,moo04} and indicates that the predictive power of the theories on food webs may be questioned if they do not include the detritus path.

At the same time, since the detrital food web is more complex than the grazing one in terms of interaction patterns, these results provide a hint 
about complexity promoting stability. This was the prevalent belief in the early days of ecology~\cite{mac55,odu53,elt58}, then it was questioned by 
May~\cite{may72} and to this date is one of the hottest topics in the ecological research.

The critical values of the predators' metabolic rate, below which the system is
driven to extinction, indicate that the predators are the key
species in the system, in agreement with~\cite{est11}.

Analysis of the time evolution of species has shown that in the grazing model the amplitudes of the fluctuations are much larger. The presence of 
detritus ($D$), which constitutes food for the lowest level species $R$, leads to a damping of the fluctuations in the densities of species, hence it reduces the risk of extinction of some species and
 a possible collapse of the FW. Without detritus the $R$ are always and everywhere
available to $P$, hence the number of $P$ in the absence of $W$ increases very fast.
Once even a small group of predators enters in that region, they find abundant food,
they proliferate fast and that leads to equally fast decrease in the number of $P$.
Without food the number of $W$ goes down and the process can repeat itself. When $R$
needs $D$ for survival, the number of $R$ is controlled from two sides - $D$ and
$P$. Therefore the population of $P$ finds less food, its growth is more limited
than in the previous case and as a consequence, an invasion of $W$ leads to smaller fluctuations of the abundances of either $P$ or $W$. The presence 
of $D$
creates a feedback mechanism which damps the oscillations in the number of species, making thus a FW more
stable.~\cite{sha81}.

\section*{Acknowledgements}
Thanks to Dariusz Grech for useful discussions on different aspects of time series and to two anonymous referees for their constructive suggestions.

\end{document}